\documentclass[aps,preprint,nofootinbib,amsmath,amssymb]{revtex4}
\usepackage{graphicx}
\usepackage{color}

\def\be{\begin{eqnarray}}
\def\ed{\end{eqnarray}}
\def\non{\nonumber}
\def\UP{\cal U}
\def\g5{\gamma_{5}}
\def\ga{\gamma}
\def\la{\langle}
\def\ra{\rangle}
\def\e{\epsilon}
\def\BZ{\cal BZ}

\def\GeV{\textrm{GeV}}
\def\TeV{\textrm{TeV}}

\def\ASM{A^{SM}}
\def\AUP{A^{\UP}}

\def\BR{\mathcal{B}}
\def\ACP{\mathcal{A}_{CP}}

\def\pP{P^{\prime}}
\def\pT{T^{\prime}}
\def\pC{C^{\prime}}
\def\pPEW{P^{\prime}_{EW}}

\def\PEW{P_{EW}}

\begin{document}
\title{ \bf
Investigation of $B_{u,d}\to (\pi,\, K) \pi$ decays  within unparticle physics}

\vspace{5cm}

\author{
Chuan-Hung Chen$^{1}$\footnote{E-mail: physchen@mail.ncku.edu.tw },~~
C. S. Kim$^{2}$\footnote{E-mail: cskim@yonsei.ac.kr},~~
Yeo Woong Yoon$^{2}$\footnote{E-mail: ywyoon@yonsei.ac.kr}
}
\address{
$^{1}$ Department of Physics, National Cheng-Kung University, Tainan 701, Taiwan \\
$^{2}$ Department of Physics, Yonsei University, Seoul 120-479, Korea
}

\begin{abstract}
\noindent We investigate the implication of unparticle physics on
the $B_{u,d}\to (\pi,\, K) \pi$ decays under the constraints of the
$B_{d,s}-\bar B_{d,s}$ mixing. We found that not only the unparticle
parameters that belong to the flavor changing neutral current (FCNC)
processes but also scaling dimension  $d_{\UP}$ could be constrained
by the $B_{d,s}-\bar B_{d,s}$ mixing phenomenology. Employing the
minimum $\chi^2$ analysis to the $B_{u,d}\to (\pi,\, K) \pi$ decays
with the constraints of $B_{d,s}$ mixing, we find that the puzzle of
large branching ratio for $B_{d}\to \pi^0 \pi^0$ and the discrepancy
between the standard model estimation and data for the direct CP
asymmetry of $B^+ \to K^+ \pi^0$ and $B_d \to \pi^+\pi^-$ can be
resolved well. However, the mixing induced CP asymmetry of $B_d\to
K_S \pi^0$ could not be well accommodated by the unparticle
contributions.
\\
\end{abstract}

\maketitle

\newpage
\section{Introduction}

Recently some incomprehensible phenomena at $B$ factories have been
explored, especially $B_{u,d}\to
(\pi,\, K)\pi$ decays.
 Firstly, the observations on the large branching
ratio (BR) for $B_d\to \pi^0 \pi^0$ decay with the world average
${\cal B}(B_d\to \pi^0 \pi^0)=(1.31 \pm 0.21)\times 10^{-6}$ and the
direct CP asymmetry for $B_{d}\to \pi^{+} \pi^{-}$ with
$\ACP(B_d\to \pi^+ \pi^{-})=0.38\pm 0.07$ \cite{HFAG} are
inconsistent with  the theoretical estimations of around $ 0.5\times
10^{-6}$ and $10-20\%$, respectively.
Secondly, a disagreement in the  CP asymmetries (CPAs) for
$B_{d}\to K^+ \pi^-$ and $B^{+}\to K^+ \pi^0 $ has been
observed to be $-0.097\pm 0.012$ and $0.050\pm 0.025$ \cite{HFAG},
respectively, while the naive estimation is $\Delta_{CP}\equiv
\ACP(B^{+}\to K^+ \pi^0)-\ACP(B_{d}\to K^+ \pi^-)\sim 0$.
Although many theoretical calculations based on QCDF \cite{QCDF},
PQCD \cite{PQCD1} and SCET \cite{SCET} have been tried to
produce the consistencies with data in the framework of standard
model (SM), however, the results have not been conclusive yet
\cite{JRS}. For instance, the recent PQCD result for the
$\Delta_{CP}$ is $0.08\pm0.09$, which is actually consistent with
the data. However, the PQCD prediction $\ACP (B^+ \to K^+
\pi^0)_{PQCD} = -0.01^{+0.03}_{-0.05}$ still has $1.4\sigma$
difference from the current experimental data \cite{Li:2005kt}.
In addition, the difference between $(\sin2\beta)_{K_S \pi^0}$ and
$(\sin2\beta)_{J/\Psi K_S}$ in the mixing-induced CPA from the PQCD
prediction is $0.065\pm 0.04$, which shows about $2\sigma$ off the
data $-0.30\pm 0.19$. Hence, the inconsistencies between data and
theoretical predictions provide a strong indication to investigate
the new physics beyond SM.

%****  Introduction to Unparticles  *******
%*********************************************

There introduced many extensions of the SM, and enormous studies
have been done on searching some specific models beyond SM, e.g. on
supersymmetric model \cite{SUSY}, extra-dimension model
\cite{EXTRADIM}, left-right symmetric model \cite{LRmodel} and
flavor-changing $Z'$ model \cite{Zprime}. Although new physics
effects will be introduced, however, in phenomenological sense we
just bring more particles and their related interactions to our
system. Recently, Georgi proposed completely different stuff and
suggested that an invisible sector, dictated by the scale invariance
and coupled weakly to the particles of the SM, may exist in our
universe \cite{Georgi1, Georgi2}. Unlike the concept of particles in
the SM or its normal extensions where the particles own the definite
mass, the scale invariant stuff cannot have a definite mass unless
it is zero. Therefore, if the peculiar stuff exists, it should be
made of {\it unparticles} \cite{Georgi1}. Furthermore, in terms of
the two-point function with the scale invariance, it is found that
the unparticle with the scaling dimension $d_{\UP}$ behaves like a
non-integral number $d_{\UP}$ of invisible particles \cite{Georgi1}.
Based on Georgi's proposal, the phenomenology of unparticle physics
has been extensively studied in Refs.
\cite{Georgi1,Georgi2,CKY,CH12,UNPARTICLE,Mohanta:2007ad}. For
illustration, some examples such as $t\to u+ \UP$ and $e^{+}
e^{-}\to \mu^{+} \mu^{-}$ have been introduced to display the
unparticle properties. In addition, it is also suggested that the
unparticle production in high energy colliders might be detected by
searching for the missing energy and momentum distributions
\cite{Georgi1,Georgi2,CKY}. Nevertheless, we have to point out that
flavor factories with high luminosities, such as SuperKEKB
\cite{SuperKEKB}, SuperB \cite{SuperB} and LHCb \cite{LHCb} etc,
should also provide good environments to search for the unparticle
effects in indirect way.

Besides the weird property of non-integral number of unparticles, the
most astonished effect is that an unparticle could carry a peculiar
CP conserving phase associated with its propagator in the time-like
region \cite{Georgi2,CKY}. It has been pointed out that the unparticle
phase plays a role like a strong phase and has an important impact
on direct CP violation (CPV) \cite{CH12}. In this paper, we will
make detailed analysis to examine whether the puzzles in $B_{u,d}\to
(\pi,\, K)\pi$ decays with the $B_{d,s}-\bar B_{d,s}$ mixing constraints
could be resolved when the invisible unparticle
stuff is introduced to the SM.

In order to study the flavor physics associated with scale invariant
stuff, we follow the scheme proposed in Ref.~\cite{Georgi1}. For the
system with the scale invariance, there exist so-called Banks-Zaks
($\cal BZ$) fields that have a nontrivial infrared fixed point at a
very high energy scale \cite{BZ}. Subsequently, with the dimensional
transmutation at the $\Lambda_{\UP}$ scale, the $\BZ$ operators
composed of $\BZ$ fields will match onto unparticle operators. We
consider only vector unparticle operator in the following analysis.
Then, the effective interactions for unparticle stuff and the particles of
the SM are adopted to be
\be%
\frac{C^{q'q}_L}{\Lambda^{d_{\UP}-1}_{\UP}} \bar q'\ga_{\mu}
(1-\ga_5)q{\cal O}^{\mu}_{\UP} +
\frac{C^{q'q}_R}{\Lambda^{d_{\UP}-1}_{\UP}} \bar q'\ga_{\mu}
(1+\ga_5)q{\cal O}^{\mu}_{\UP}\,, \label{eq:int_UP}
\ed%
where $C^{q'q}_{L, R}$ are effective coefficient functions and
${\cal O}^{\mu}_{\UP}$ denotes the spin-1 unparticle operator with
scaling dimension $d_{\UP}$ and is assumed to be hermitian and
transverse $\partial_{\mu} {\cal O}^{\mu}_{\UP}=0$. Since so far the
theory for $\BZ$ fields and their interactions with SM particles is
uncertain, here $C^{q'q}_{L, R}$ are regarded as free parameters.
With scale invariance, the propagator of vector unparticle can be
obtained by \cite{Georgi2,CKY}
 \be%
 && \int d^4x e^{i p\cdot x} \la 0| T\left( O^{\mu}_{\UP}(x)
O^{\nu}_{\UP}(0)\right) |0 \ra \non\\
&=& i\Delta_{\UP}(p^2)\left(-g^{\mu \nu} + \frac{
p^{\mu}p^{\nu}}{p^2}\right)\,e^{-i\phi_{\UP}}\,, \label{eq:uprop}
 \ed
 with $\phi_{\UP}=(d_{\UP}-2)\pi$ and
 \be%
 \label{PAR-UP}
    \Delta_{\UP}(p^2)&=& \frac{A_{d_{\UP}}}{2\sin(d_{\UP}
\pi)} \frac{ 1}{\left(p^2
+i\e\right)^{2-d_{\UP}}}\,, \non\\
A_{d_{\UP}}&=&  \frac{16 \pi^{5/2}}{(2\pi)^{2d_{\UP}}}
\frac{\Gamma(d_{\UP}+1/2)}{\Gamma(d_{\UP}-1) \Gamma(2 d_{\UP})}\,,
\label{eq:pro_up}
\ed%
where $\phi_{\UP}$ could be regarded as a CP conserving phase
\cite{Georgi2,CKY,CH12}. We note that when unparticle stuff is
realized in the framework of conformal field theories, the
propagators for vector and tensor unparticles should be modified
\cite{Grinstein:2008qk}. Although conformal invariance typically
implies scale invariance, however, in principle it is not necessary.
Unparticle stuff with scalar invariance in 2D spacetime has been
investigated in Ref.~\cite{Georgi3}. Hence, in this work, we still
concentrate on the stuff built out of only scale invariance. For
simplicity, we set the unknown scale factor $\Lambda_{\UP}$ to be $1
\,\TeV$ throughout the analysis.

\section{Constraints of $B_{d,s}-\bar B_{d,s}$ mixing }
Since the tree level FCNC processes are allowed in
the unparticle physics, it is expected that the $B_{d,s}-\bar B_{d,s}$ mixings
offer strong constraints on the unparticle parameters of $C^{db}_L, C^{db}_R$ and
$C^{sb}_L, C^{sb}_R$. Moreover, it will be shown that the scaling
dimension $d_{\UP}$ also can be constrained by $B_{d,s}-\bar B_{d,s}$ mixings.

First of all, we separate the matrix element of $\Delta B =2$ transition for the
$B_{q}^0-\bar B_{q}^0$ mixing $(q=d,s)$, which is denoted by $M^q_{12}$, into
the SM and unparticle contribution as follows.
 \be
 M^{q}_{12}=M^{q,SM}_{12}+M^{q,NP}_{12}=|M^{q,SM}_{12}|e^{i\phi^{SM}_{q}}
 +|M^{q,NP}_{12}|e^{i\phi^{NP}_{q}}\,. \label{eq:mq12}
 \ed
 where the $\phi_q^{SM}$ and $\phi_q^{NP}$ represent the phases of mixing
 amplitudes. For the second term, we use the superscript {\it `NP'} in order to
 represent general new physics (NP) contribution. Later on, we regard this
 $M^{q,NP}_{12}$ as the unparticle mixing amplitude.

 As is well known, the magnitude of total mixing amplitude $|M^{q}_{12}|$
 is given by the $B^0_q-\bar B^0_q$ oscillating frequency as follows
\begin{eqnarray}
\Delta M_q = 2 | M^q_{12} |\,.
\end{eqnarray}
And the mixing phase $\phi_q \equiv \arg M^q_{12}$ can be obtained from the
mixing induced CP asymmetry of $b\to c \bar c s$ processes. We summarize
current experimental data in Table~\ref{tab:exp}.
%%%%%%%%%%%%%%%%%%%%%%%%%%%%%%%%%%%%%%%%%%%%%%%%%%%%%%%%%%%%%%%%%%%%%%%%%%%%%%%%%
%%%%%%%%%%%%%%%%%%%%%%%%%%% TABLE exp %%%%%%%%%%%%%%%%%%%%%%%%%%%%%%%%%%%%%%%%%%%
\begin{table}[t]
\caption{ Experimental values for the $B_{d,s}-\bar B_{d,s}$ mixings.
Even though D0 collaboration recently have measured $\phi_s$ \cite{Walder:2007pr},
the data is not used in our analysis because of huge error of it.}
\label{tab:exp}
\begin{tabular}{ccc}
\hline
\hline
~~ observables ~~&~~  values  ~~&~~ note\\
\hline
$\Delta M_d$ ~~&~~ $(~0.507 \pm 0.004~)~ \textrm{ps}^{-1} $ ~~&~~HFAG \cite{HFAG}\\
$\Delta M_s$ ~~&~~ $(~17.77\pm0.12~)~\textrm{ps}^{-1}  $ ~~&~~
CDF \cite{Abulencia:2006ze}\\
$\phi_d$  ~~&~~  $43^\circ \pm 2^\circ$  ~~&~~ HFAG \cite{HFAG} \\
\hline
\hline
\end{tabular}
\begin{tabular}{l}
\end{tabular}
\end{table}
%%%%%%%%%%%%%%%%%%%%%%%%%%%%%%%%%%%%%%%%%%%%%%%%%%%%%%%%%%%%%%%%%%%%%%%%%%%%%%%%%
%%%%%%%%%%%%%%%%%%%%%%%%%%%%%%%%%%%%%%%%%%%%%%%%%%%%%%%%%%%%%%%%%%%%%%%%%%%%%%%%%

 The SM mixing amplitude reads
\begin{eqnarray}
M^{q,SM}_{12}= \frac{G_F^2 m_W^2}{12\pi^2} m_{B_q} f_{B_q}^2 \hat
B_{B_q} (V_{tq}^* V_{tb})^2 \hat \eta^B S_0(x_t)\,,
\label{eq:mixing_sm}
\end{eqnarray}
where $\hat \eta^B = 0.552$ is short distance QCD correction term
\cite{Buchalla:1995vs}, and $ S_0(x_t)=2.35\pm 0.06$ is an Inami-Lim
function for the t-quark exchange in the loop diagram
\cite{Inami:1980fz}. The quantities of $f_{B_q}$ and $B_{B_q}$ are
non-perturbative parameters which can be obtained from the lattice
calculations. We follow the procedure given in Ref. \cite{Ball:2006xx} for
dealing with these non-perturbative parameters. The procedure mainly employs the
result of lattice calculations in two different ways. The one is to use the
result of JLQCD collaboration \cite{Aoki:2003xb}, and the other is to combine
the results of JLQCD and HPQCD \cite{Gray:2005ad} collaborations.
We note that one can obtain the SM mixing phases from Eq.(\ref{eq:mixing_sm})
as follows.
\begin{eqnarray}
\label{PHIqSM}
\phi^{SM}_d = 2 \beta, ~~~~~ \phi^{SM}_s = -2 \lambda^2 \eta \,,
\end{eqnarray}
where $\beta$ is an angle of CKM unitarity triangle, $\lambda$ and $\eta$ are
from the Wolfenstein parametrization \cite{Wolfenstein:1983yz}.
We use the result of UTfit \cite{UTfit} from the tree level processes for the
 $\beta , R_t$ and $\bar \eta$, where $R_t \equiv \sqrt{ (1-\bar\rho)^2 +
\bar \eta^2}$ and ($\bar \rho,\bar \eta)$ is the apex of the CKM
unitary triangle. For the $|V_{cb}|$, we adopt the result of global fit to
moment of inclusive distributions in $B \to X_c l \nu_l$, which is  performed
in the framework of heavy quark expansions with kinetic scheme
\cite{Buchmuller:2005zv}.
After putting all the SM input parameters into Eq. (\ref{eq:mixing_sm}),
the SM mixing amplitudes are obtained. And using the experimental data
shown in the Table~\ref{tab:exp}, the NP mixing amplitudes for the $B_d-\bar B_d$ mixing
could be gained through the Eq. (\ref{eq:mq12}).
The numerical values are summarized in Table~\ref{tab:para}.
%%%%%%%%%%%%%%%%%%%%%%%%%%%%%%%%%%%%%%%%%%%%%%%%%%%%%%%%%%%%%%%%%%%%%%
%%%%%%%%%%%%%%%%%%%%%%%% TABLE tab:para %%%%%%%%%%%%%%%%%%%%%%%%%%%%%%
\begin{table}[h]
\caption{ Numerical values for the SM $B_{d,s}-\bar B_{d,s}$ mixing amplitudes.
The values of the NP $B_d-\bar B_d$ mixing amplitude are obtained from the
experimental data and the Eq. (\ref{eq:mq12}). The case (a) denotes JLQCD, while
the case (b) denotes (HP+JL)QCD. $M_{12}^{q,NP} = M_{12}^{q,\UP}$ should be
considered $(q=s,d)$.}
\label{tab:para}
\begin{tabular}{cc|cc}
\hline
\hline
~ SM parameters ~&~  values  ~&~ NP  parameters ~&~ values\\
\hline
~$2|M_{12}^{d,SM}|$
~&~
$\begin{array}{l}
0.75^{+0.20}_{-0.26} ~\textrm{ps}^{-1} ~~~~~\textrm{ (a) }\\
0.97\pm0.29 ~\textrm{ps}^{-1} ~~\textrm{ (b) }
\end{array} $
~&~
$2|M_{12}^{d,NP}|$
~&~
$\begin{array}{l}
    0.25\pm0.26 ~\textrm{ps}^{-1} ~~\textrm{ (a) }\\
    0.46\pm0.29 ~\textrm{ps}^{-1} ~~\textrm{ (b) }
\end{array} $
\\
\hline
~$\phi_d^{SM}$ ~& $45.2^\circ\pm5.7^\circ$ ~&~
~$\phi_d^{NP}$ ~&~
$\begin{array}{l}
    -130^\circ \pm 180^\circ ~~~\textrm{ (a) }\\
    -132^\circ \pm 12^\circ ~~~~\textrm{ (b) }
\end{array}$  \\
\hline
~$2|M_{12}^{s,SM}|$
~&~
$\begin{array}{l}
 16.4\pm2.8 ~\textrm{ps}^{-1} ~~~~\textrm{ (a) }\\
 23.8\pm5.9 ~\textrm{ps}^{-1} ~~~~\textrm{ (b) }
\end{array} $
~&~
$2|M_{12}^{s,NP}|$
~&~
-
\\
\hline
~$\phi_s^{SM}$ ~& $-2.3^\circ \pm 0.2^\circ$ ~&~
~$\phi_s^{NP}$ ~&~ - \\
\hline
\hline
\end{tabular}
\begin{tabular}{l}
\end{tabular}
\end{table}
%%%%%%%%%%%%%%%%%%%%%%%%%%%%%%%%%%%%%%%%%%%%%%%%%%%%%%%%%%%%%%%%%%%%%%
%%%%%%%%%%%%%%%%%%%%%%%%%%%%%%%%%%%%%%%%%%%%%%%%%%%%%%%%%%%%%%%%%%%%%%

\begin{figure}[htbp]
\includegraphics*[width=3.0 in]{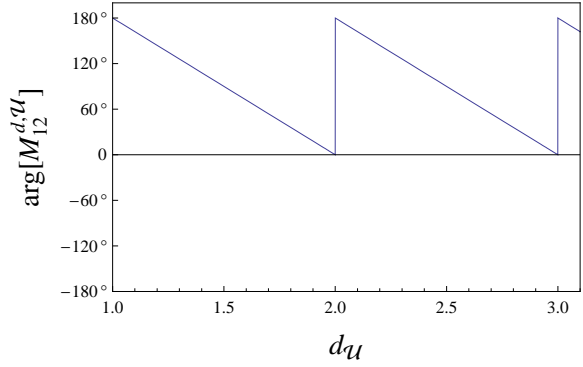}
\caption{Plot of $\arg (M_{12}^{d,\UP})$ versus $d_{\UP}$ within the
range of $(-180^\circ,180^\circ)$.}
 \label{Fig-phase}
\end{figure}

As for the unparticle contribution to the mixing amplitude, we begin
with the effective Hamiltonian for $\Delta B=2$ processes in
unparticle sector such as
 \be
{\cal H}^{q,\UP}&=& 2\cdot \frac{1}{4}\cdot
\left(\frac{p^2}{\Lambda^2_{\UP}}\right)^{d_{\UP}-1}  \frac{1}{p^2}~
\frac{A_{\UP}}{2\sin d_{\UP}\pi} e^{-i\phi_{\UP}} \non\\
&\times & \left[ -\bar q \ga_{\mu} \left(C^{qb}_{L} (1-\ga_5) +
C^{qb}_{R}(1+\ga_5) \right) b\; \bar q \ga^{\mu} \left( C^{qb}_{L}
(1-\ga_5) + C^{qb}_{R}(1+\ga_5) \right)b \right.\non\\
&+&\left. \frac{1}{p^2} \bar q \not p \left( C^{qb}_{L}(1-\ga_5) +
C^{qb}_{R} (1+\ga_5)\right)b\; \bar q \not p
\left(C^{qb}_{L}(1-\ga_5) + C^{qb}_{R}(1+\ga_5) \right) b\right]\,.
 \ed
The factor 2 is from the fact that there are $s$ and $t-$channel
which give same result, and the factor 1/4 is due to the Wick
contraction factor  \cite{Chen:2007cz}.
 From this effective Hamiltonian, the transition matrix elements can
be shown as
\begin{eqnarray}
\label{Mq12U}
M_{12}^{q,\UP} = -\frac{\Delta_{\UP}(p^2)}{\left(\Lambda^2_{\UP}
\right)^{d_{\UP}-1}}e^{-i \phi_{\UP}}
m_{B_q} f^2_{B_q} \hat
B_{B_q} a^{q,\UP}_{\textrm{mix}}\,,
\end{eqnarray}
where the $a^{q,\UP}_{\textrm{mix}}$ is defined by
\begin{eqnarray}
\label{aqU} a^{q,\UP}_{\textrm{mix}}\equiv
\left[(C^{qb}_L)^2+(C^{qb}_R)^2 \right]\left[\frac{2}{3} -
\frac{5}{12} \frac{m^2_{B_q}}{p^2} \right] +C^{qb}_{L} C^{qb}_{R}
\left[- \frac{5}{3} + \frac{7}{6}\frac{m^2_{B_q}}{p^2} \right]\,.
\end{eqnarray}
with $p^2=m^2_{B_{q}}$. $\Delta_{\UP}(p^2)$ is given in Eq. (\ref{PAR-UP}).
In order to get the constraints on the unparticle parameters from the NP
parameter regions shown in the Table~\ref{tab:para}, we first consider the phase of
$M_{12}^{q,\UP}$. It depends on the scaling dimension $d_{\UP}$ through
the $ e^{-i \phi_{\UP}}$ term and the sign of $\sin(d_{\UP}\pi)$ in the
$\Delta_{\UP}(p^2)$. The plot of $\arg(M_{12}^{d,\UP})$ versus $d_{\UP}$
is shown in Fig \ref{Fig-phase}.
%%%%%%%%%%%%%%%%%%%%%%%%%%%%%%%%%%%%%%%%%%%%%%%%%%%%%%%%%%  Fig-phase
%%%%%%%%%%%%%%%%%%%%%%%%%%%%%%%%%%%%%%%%%%%%%%%%%%%%%%%%%%

%%%%%%%%%%%%%%%%%%%%%%%%%%%%%%%%%%%%%%%%%%%%%%%%%%%%%%%%%%
%%%%%%%%%%%%%%%%%%%%%%%%%%%%%%%%%%%%%%%%%%%%%%%%%%%%%%%%%%
It is interesting that $\arg(M_{12}^{d,\UP})$ takes only positive value as
displayed in Fig. \ref{Fig-phase}. Therefore, as we can see from Table~\ref{tab:para}.
(HP+JL)QCD case can not give a solution of $d_{\UP}$
while all value of $\arg(M_{12}^{d,\UP})$ is possible in JLQCD case.
So, we take only the JLQCD case. Although we can not
determine $d_{\UP}$ in the JLQCD case presently,
if the SM prediction for the
$B_d - \bar B_d$ mixing amplitude becomes more precise in  future,
it can give strong constraint on $d_{\UP}$ from the plot of
$\arg(M_{12}^{d,\UP})$ versus $d_{\UP}$. Here, we assume
$d_{\UP}=1.5$ for the remaining analysis.
Then, the magnitude of
unparticle transition matrix element $M_{12}^{s,\UP}$ can be
obtained through  Eq. (\ref{eq:mq12}) and the values in
Table~\ref{tab:para} for the JLQCD case with the experimental data
as $2|M_{12}^{s,\UP}| = 7.6 \pm 6.6~\textrm{ps}^{-1}$. Using this value and the
value for $2|M_{12}^{d,\UP}|$ in Table~\ref{tab:para}, we can see that the
mixing parameter $a^{q,\UP}_{\textrm{mix}}$ should be strongly suppressed as follows.
\begin{eqnarray}
\label{eq:amix}
|a^{d,\UP}_{\textrm{mix}}| = (1.1 \pm 1.3) \times 10^{-8}\,, ~ ~ ~ ~
|a^{s,\UP}_{\textrm{mix}}| = (2.6 \pm 2.3) \times 10^{-7}\,.
\end{eqnarray}
And, the Eq. (\ref{aqU}) leads to
\begin{eqnarray}
\label{eq:CLCR}
|C_L^{qb} - C_R^{qb}| = 2 \sqrt{ a^{q,\UP}_{\textrm{mix}}}\,.
\end{eqnarray}
Therefore, the parameters $C^{qb}_L$ and $C^{qb}_R~(q=d,s)$ are strongly correlated
under the $B_{d,s}-\bar B_{d,s}$ mixing.
In the next section, these strong constraints on the unparticle parameters
will be used for fitting the parameters in $B_{u,d}\to (\pi,\,K) \pi$ decays

%%%%%%%%%%%%%%%%%%%%%%%%%%%%%%%%%%%%%%%%%%%%%%%%%%%%%%%%%%%%%%%%%%%%%%%%%%%%%%%%%
\section{$B_{u,d}\to (\pi,\,K) \pi$ decays and the Unparticle Contributions   }
%%%%%%%%%%%%%%%%%%%%%%%%%%%%%%%%%%%%%%%%%%%%%%%%%%%%%%%%%%%%%%%%%%%%%%%%%%%%%%%%%
 According to the Eqs.~(\ref{eq:int_UP}) and (\ref{eq:uprop}),
 the effective Hamiltonian for $b\to q \bar q' q'$ decays is obtained by
\be%
{\cal H}_{\UP}&=& - C_{\UP}(q^2)  \left( C^{qb}_L (\bar q b)_{V-A}+
C^{qb}_{R}(\bar q b)_{V+A}\right) \left( C^{q'q'}_L (\bar q'
q')_{V-A}+ C^{q'q'}_{R}(\bar q'q')_{V+A}\right)\,, \label{eq:effH}
\ed%
where $q=(d,\, s)$, $q'=(u,\,d,\,s,\,c)$, $(\bar f' f)_{V\pm A}=\bar
f' \ga_{\mu}(1\pm \ga_5) f$ and
 \be
C_{\UP}(q^2)=\frac{\Delta_{\UP}(q^2)}{\left(\Lambda^2_{\UP}
\right)^{d_{\UP}-1}}e^{-i \phi_{\UP}} \,.
 \ed
Based on this effective Hamiltonian, we study the unparticle
contributions to the decay amplitudes for $B_{u,d}\to (\pi,\,K)
\pi$.  It is known that the most uncertain theoretical calculations
for two-body exclusive decays are the QCD hadronic transition matrix
elements. To deal with the hadronic matrix elements, we adopt recent
perturbative QCD (PQCD) calculations for the SM amplitudes and naive
factorization (NF) approach for the amplitudes of unparticle
contributions. The SM amplitudes for $B_{u,d}\to (\pi,\, K) \pi$
decays can be parameterized in the context of quark diagram approach
(QDA) \cite{Gronau:1994bn} as follows
\begin{eqnarray}
\label{amp-pipi+0} \sqrt{2}\ASM(B^+ \to \pi^+ \pi^0) &=&
-T e^{i\gamma} -C e^{i\gamma} -\PEW e^{-i \beta} , \\
\label{amp-pipi+-} \ASM(B_d \to \pi^+ \pi^-) &=&
-T e^{i\gamma} - P e^{-i \beta}, \\
\label{amp-pipi00} \sqrt{2} \ASM(B_d \to \pi^0 \pi^0) &=& -C
e^{i\gamma} + P e^{-i \beta} -\PEW e^{-i \beta}, \\
\label{amp0+} \ASM(B^+ \to K^0 \pi^+) &=& \pP, \\
\label{amp+-} \ASM(B_d \to K^+ \pi^-) &=& -\pP
  - \pT e^{i \gamma} , \\
\label{amp+0} \sqrt{2} \ASM(B^+ \to K^+ \pi^0) &=& -\pP - \pT e^{i
\gamma}
 - \pC e^{i \gamma}  - \pPEW , \\
\label{amp00} \sqrt{2} \ASM(B_d \to K^0 \pi^0) &=& \pP - \pC e^{i
\gamma}  - \pPEW \,,
\end{eqnarray}
where $T^{(\prime)}$ and $C^{(\prime)}$ denote the tree
color-allowed and -suppressed amplitudes for $B_{u,d}\to \pi (K) \pi$,
respectively, while $P^{(\prime)} (P^{(\prime)}_{EW})$ is gluonic
(electroweak) penguin amplitude. All CP-conserving phases are
included in these parameters. The phase $\gamma(\beta)$ is the CP
violating phase in the SM and from $V_{ub}(V_{td})$. Table~\ref{tab:PQCD} shows
the recent PQCD result for the values of each topological
parameters \cite{Li:2005kt,Lielal}.
%%%%%%%%%%%%%%%%%%%%%%%%%%%%%%%%%%%%%%%%%%%%%%%%%%%%%%%%%%%%%%%%%
%%%%%%%%%%%%%%%%%%%%%  TABLE PQCD %%%%%%%%%%%%%%%%%%%%%%%%%%%%%%%%
\begin{table}[h]
\caption{Recent PQCD predictions for the topological parameters of
$B_{u,d}\to (\pi,\,K) \pi$ in unit of $10^{-5}$ GeV
. The phases are indicating strong phases of the
parameters in radian unit. The predictions include NLO calculation.}
\label{tab:PQCD}
\begin{tabular}{c|cc | c|cc}
\hline
\hline
Topology   ~~&~~ Abs  ~~&~~ Arg ~~&~~ Topology ~~&~~ Abs ~~&~~ Arg\\
\hline
$\pP$ ~~&~~ $43.6^{+10.8}_{-8.0}$ ~&~~ $2.9^{+0.1}_{-0.2}$  ~~&~~ $T$  ~~&~~  $23.2^{+8.0}_{-6.1}$~&~~ $0.0\pm 0.0$\\
$\pT$ ~~&~~ $ 6.5^{+2.4}_{-1.8}$  ~&~~ $0.1\pm 0.0$ ~~&~~ $P$  ~~&~~  $ 5.6^{+1.2}_{-0.8}$~&~~ $-0.4^{+0.2}_{-0.1}$\\
$\pPEW$ ~~&~~ $ 5.4^{+1.4}_{-1.0}$ ~&~~ $-1.3\pm 0.1$  ~~&~~ $C$  ~~&~~  $ 4.3^{+2.1}_{-1.5}$~&~~ $-1.1\pm 0.0$\\
$\pC$ ~~&~~ $ 1.7^{+0.9}_{-0.6} $  ~&~~ $-3.0\pm 0.0$ ~~&~~ $P_{EW}$  ~~&~~  $ 0.7^{+0.1}_{-0.1}$~&~~ $-0.1\pm 0.0$\\
\hline
\hline
\end{tabular}
\end{table}
%%%%%%%%%%%%%%%%%%%%%%%%%%%%%%%%%%%%%%%%%%%%%%%%%%%%%%%%%%%%%%%%%%%%%%
%%%%%%%%%%%%%%%%%%%%%%%%%%%%%%%%%%%%%%%%%%%%%%%%%%%%%%%%%%%%%%%%%%%%%%

\begin{figure}[htbp]
\includegraphics*[width=4.0 in]{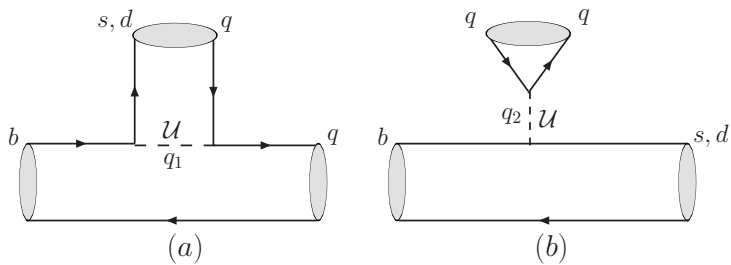}
\caption{Typical flavor diagrams for $B\to (\pi, K) \pi $ decays
mediated by unparticle with q=u and d, where $q_{1,2}$ are the
momenta of unparticle.}
 \label{fig:BpiK_un}
\end{figure}

For deriving the unparticle contributions, the definitions for
relevant decay constants and form factors are given by
\be%
\la P(p) |\bar q \ga_{\mu} \ga_5 u|0\ra&=& if_{P} p_{\mu}\,,\non\\
\la P(p)|\bar q \ga_{5} u|0\ra&=& if_{P}
m^0_{P} \,,\non\\
 \la P(p) | \bar q \ga_{\mu} b|\bar B(p_B)\ra&=&
\left[(p_B+p)_{\mu}-{m^2_B\over q^2}\,
q_{\mu}\right]F^{BP}_{1}(q^2)+ {m^2_B\over q^2}\, q_{\mu}
F^{BP}_0(q^2)\,,
\ed%
with $P=(\pi,\, K)$, $q=p_B-p$ and $m^0_P=m^2_P/(m_q+m_u)$. Here,
due to $m_{P}\ll m_B$, we have neglected the $m^2_{P}$ effects in
$B\to P$ transition matrix element. Subsequently, by
considering various flavor diagrams in which the typical diagrams
mediated by unparticle are illustrated in Fig. \ref{fig:BpiK_un},
%%%%%%%%%%%%%%%%%%%%%%%%%%%%%%%%%%%%%%%%%%%%%%%%%%%%%%%%%%%%%
%%%%%%%%%%%%%%%%%%%%%%%%%%%%%%%%%%%%%%%%%%%%%%%%%%%%%%%%%%%%
the unparticle amplitudes for $B_{u,d}\to (\pi,\, K)\pi$ decays
within NF approach are obtained to be
\begin{eqnarray}
A^{\UP}(B \to \pi^{i} \pi^{j}) &=& C_{\UP}(q_1^2) f_{\pi} m^2_{B}
F^{B\pi}_{0}(m^2_{\pi})  a_{\textrm{dec}}^{{\UP}, \pi^{i} \pi^{j}}\,,
\\
A^{\UP}(B \to K^{i} \pi^{j}) &=&  C_{\UP}(q_1^2) f_{K} m^2_{B}
F^{B\pi}_{0}(m^2_{K}) a_{\textrm{dec}}^{{\UP}, K^{i} \pi^{j}}\,,
\end{eqnarray}
where the coefficients $a_{\textrm{dec}}^{\UP}$s are defined in
Table~\ref{tab:amp_up},
%%%%%%%%%%%%%%%%%%%%%%%%%%%%%%%%%%%%%%%%%%%%%%%%%%%%%%%%%%%%%%%%%%%%%%
%%%%%%%%%%%%%%%%%%%%%%%%%%  TABLE amp_up %%%%%%%%%%%%%%%%%%%%%%%%%%%%%%
\begin{table}[t]
\caption{ The definition of coefficients for the unparticle
amplitudes in $B_{u,d}\to (\pi,\,K) \pi$ decays within NF
approach.}\label{tab:amp_up}
\begin{tabular}{c|c}
\hline
~ decay mode  ~&  $a^{\UP}_{\textrm{dec}}$ \\
\hline
\hline
$\pi^{+}\pi^{-}$   &
$-\frac{1}{N_c}\Big( \left(C^{db}_{L} C^{uu}_{L}-C^{db}_{R}C^{uu}_{R} \right)
 + 2 r^{\pi}_1\left( C^{db}_{L} C^{uu}_{R}-C^{db}_{R}C^{uu}_{L} \right)\Big)$ \\
\hline
$\pi^{+}\pi^{0}$  &
$\begin{array}{c}
-\frac{1}{\sqrt{2}N_c}\Big(
 \left(C^{db}_{L} (C^{uu}_{L}-C^{dd}_{L})-C^{db}_{R}(C^{uu}_{R}-C^{dd}_{R}) \right) \\
 + 2 r^{\pi}_2\left( C^{db}_{L}
(C^{uu}_{R}-C^{dd}_{R})-C^{db}_{R}(C^{uu}_{L}-C^{dd}_{L}) \right) \Big)\\
- \frac{C_{\UP}(q^2_{2})}{\sqrt{2} C_{\UP}(q^2_{1})}\left(C^{db}_L + C^{db}_R \right)
 \left( C^{uu}_{L}-C^{dd}_{L}-C^{uu}_{R} + C^{dd}_{R}\right) \end{array}$ \\
\hline
$\pi^{0}\pi^{0}$  &
$\begin{array}{c}
\frac{1}{\sqrt{2}N_c} \Big(
 \left(C^{db}_{L} C^{dd}_{L}-C^{db}_{R}C^{dd}_{R} \right)
 + 2 r^{\pi}_2\left( C^{db}_{L} C^{dd}_{R}-C^{db}_{R}C^{dd}_{L} \right)
 \Big) \\
 -  \frac{C_{\UP}(q^2_{2})}{\sqrt{2}C_{\UP}(q^2_{1})} \left(C^{db}_L + C^{db}_R \right)
 \left( C^{uu}_L-C^{dd}_L-C^{uu}_{R} + C^{dd}_{R}\right) \end{array} $ \\
\hline
$K^{0}\pi^{-}$   &
$\frac{1}{N_c} \Big(
 \left(C^{sb}_{L} C^{dd}_{L}-C^{sb}_{R}C^{dd}_{R} \right)
 + 2 r^{K}_1\left( C^{sb}_{L} C^{dd}_{R}-C^{sb}_{R}C^{dd}_{L} \right)
 \Big) $
\\
\hline
$K^{+}\pi^{-}$  &
$- \frac{1}{N_c} \Big(
 \left(C^{sb}_{L} C^{uu}_{L}-C^{sb}_{R}C^{uu}_{R} \right)
 + 2 r^{K}_1\left( C^{sb}_{L} C^{uu}_{R}-C^{sb}_{R}C^{uu}_{L} \right)
 \Big) $\\
\hline
$K^{+}\pi^{0}$  &
$ \begin{array}{c}
-\frac{1}{\sqrt{2}N_c} \Big( \left(C^{sb}_{L} C^{uu}_{L}-C^{sb}_{R}C^{uu}_{R} \right)
 +  2 r^{K}_1\left( C^{sb}_{L} C^{uu}_{R}-C^{sb}_{R}C^{uu}_{L}
 \right) \Big) \\
 - \frac{C_{\UP}(q^2_{2})}{\sqrt{2}C_{\UP}(q^2_{1})}
 \frac{f_{\pi}}{f_K}\frac{F^{BK}_0(m_{\pi}^2)}{F^{B\pi}_0(m_{K}^2)}
 \left(C^{sb}_L + C^{sb}_R  \right)\left( C^{uu}_{L}-C^{dd}_{L}-C^{uu}_{R} + C^{dd}_{R}\right) \end{array} $ \\
\hline
$K^{0}\pi^{0}$ &
$ \begin{array}{c}
\frac{1}{\sqrt{2}N_c} \Big( \left(
 C^{sb}_{L} C^{dd}_{L}-C^{sb}_{R}C^{dd}_{R} \right)  + 2 r^{K}_2\left( C^{sb}_{L} C^{dd}_{R}-C^{sb}_{R}C^{dd}_{L} \right)
 \Big) \\
 - \frac{C_{\UP}(q^2_{2})}{\sqrt{2}C_{\UP}(q^2_{1})}
 \frac{f_{\pi}}{f_K}\frac{F^{BK}_0(m_{\pi}^2)}{F^{B\pi}_0(m_{K}^2)}
 \left(C^{sb}_L + C^{sb}_R \right)
 \left( C^{uu}_{L}-C^{dd}_{L}-C^{uu}_{R} + C^{dd}_{R}\right) \end{array} $ \\
\hline
\end{tabular}
\end{table}
%%%%%%%%%%%%%%%%%%%%%%%%%%%%%% TABLE amp_up  %%%%%%%%%%%%%%%%%%%%%%%%%%%%%%%%%
%%%%%%%%%%%%%%%%%%%%%%%%%%%%%%%%%%%%%%%%%%%%%%%%%%%%%%%%%%%%%%%%%%%%%%%%%%
$N_c=3$ is the number of colors, $q_2^2 = m_\pi^2$, and the chiral
enhanced factor $r^\pi_{(1,2)}$ and $r^K_{(1,2)}$ are defined by
 \begin{eqnarray}
 \label{rPI}
 r^\pi_1 = \frac{m_\pi^2}{m_b (m_u + m_d)}\,, ~~~~~
 r^\pi_2 = \frac{m_\pi^2}{m_b (m_d + m_d)}\,, \non\\
 r^K_1 = \frac{m_K^2}{m_b (m_u + m_s)}\,, ~~~~~
 r^K_2 = \frac{m_K^2}{m_b (m_d + m_s)}\,.
 \end{eqnarray}
 We note that since $q_1$ in Fig.~\ref{fig:BpiK_un}(a)
involves different mesons, the estimation of $q^2_1$ should have
ambiguity. To understand the typical value of $q^2_1$, we write the
$q_1=p_B-k_2-k_3$ with $k_{2,3}$ being the momenta of valence quarks
inside the light mesons. In terms of momentum fraction of valence
quark and light-cone coordinates and by neglecting the transverse
momentum, one can get $k_{2}=(0,m_B x_{2}/\sqrt{2},\vec{0}_{\perp})$
and $k_{3}=(m_B x_{3}/\sqrt{2},0,\vec{0}_{\perp})$. As a result, we
have $q^2_1 = m^2_B(1-x_2)(1-x_3)$.
According to the behavior of leading twist wave function of light
meson, $\Phi^{\rm tw-2}\propto x(1-x)$ which is calculated by QCD
sum rules \cite{BZ_PLB633}, it is known that the maxima of $x_{2,3}$
occur at $x_2= x_3\sim 1/2$. Therefore, for numerical estimations,
the momentum transfer could be roughly taken as $q^2_1\approx
m^2_B/4$ with $m_B = 5.28$ GeV.

We discard irrelevant factor $i$ in the NF and match the
sign with the QDA parametrization; then, the total amplitude is
 \begin{equation}
 A(B \to f) = \kappa_f \ASM(B\to f) + \AUP(B \to f)\,.
 \end{equation}
Here, the $\kappa_f$ is the ratio of phase space factor coming from
the difference of notation of decay amplitude between NF and PQCD
group and is defined by
\begin{equation}
\kappa_f \equiv \sqrt{ \left( \frac{G_F^2 m_b^3}{128\pi} \right)
\big / \left( \frac{ p_f}{8\pi m_B^2} \right)} =
1.15\times 10^{-4}\,.
\end{equation}
{}From Table~\ref{tab:amp_up}, we see that besides $d_{\UP}$ and
$\Lambda_{\UP}$, the introduced new free parameters are
\begin{eqnarray}
\label{PARS}
 && C^{db}_{L}, \ \ \  C^{db}_{R},\ \ \ C^{sb}_{L},\ \ \
C^{sb}_{R}\,,
\non\\
 && C^{uu}_{L}, \ \ \  C^{uu}_{R},\ \ \ C^{dd}_{L},\ \ \
C^{dd}_{R}\,,
\end{eqnarray}
which denote the couplings of unparticle to SM particles.
First four parameters are strongly correlated by $B_{d,s}-\bar B_{d,s}$
mixing phenomena as shown in Eq.(\ref{eq:CLCR}).
If we regard all these parameters to be real number and set $\Lambda_{\UP}=1$
TeV and $d_{\UP}=1.5$ as we do in the $B_{d,s}-\bar B_{d,s}$ analysis,
8 free parameters are involved for
$B_{u,d}\to (\pi,\, K) \pi$ in the unparticle physics.

In order to fit to the data for the $B_{u,d}\to (\pi,\, K) \pi$ decays
with these 8 parameters, we perform the minimum $\chi^2$ analysis.
According to current experimental observations,  the amount of
available data for $B_{u,d}\to (\pi,\, K)\pi$ decays is $17$ as
their world averages are displayed in Table~\ref{tab:fit}
Besides the BRs, the
important quantities to display the new physics effects are the
direct and mixing induced CPAs, where they could be briefly defined
through
 \be
{\cal A}_{f} \equiv \frac{|\lambda_f|^2-1}{1+|\lambda_f|^2} \,, \ \
\ S_f \equiv \frac{2 {\rm Im}(\lambda_f)}{1+|\lambda_f|^2} \,,
 \ed
with $\lambda_f=e^{i\phi_d} \bar A_{\bar f}/A_{f}$, respectively.
For direct CPA, $f$ could be any possible final states; however, for
mixing induced CPA, $f$ could only be CP eigenstates.
Since the PQCD prediction for the SM decay amplitudes has sizable error
as shown in Table~\ref{tab:PQCD}, it should be considered for the minimum
$\chi^2$ analysis.
%%%%%%%%%%%%%%%%%%%%%%%%%%%%%%%%%%%%%%%%%%%%%%%%%%%%%%%%%%%%%%%%%%%%%%%%%%%
%%%%%%%%%%%%%%%%%%%%%%%%  TABLE fit   %%%%%%%%%%%%%%%%%%%%%%%%%%%%%%%%%%%%%
\begin{table}[pt]
\caption{ The experimental data for $B_{u,d}\to (\pi,\, K)\pi$ decays and
comparison between the experimental data and the theoretical predictions
with and without unparticle contribution. The BRs are order of $10^{-6}$.
The data is updated by September 2007. `w/o' means `without unparticle contribution'.
$\chi^2$ contributions of each
observables are shown.}
\label{tab:fit}
\begin{tabular}{c|ccc|cc}
\hline
~~~~ observables ~&~ data ~&~ theory (w/o)~&~~theory
~~&~ $\chi^2$ (w/o) ~&~ $\chi^2$  ~~~\\
\hline
$\BR(K^0 \pi^+)$ ~&~~$23.1\pm1.0$ ~&~ $23.5\pm12  $~&~ $23.1\pm11  $ ~&~~0.001 ~&~~0.0\\
$\BR(K^+ \pi^0)$ ~&~~$12.9\pm0.6$ ~&~ $13.0\pm6.2 $~&~ $12.7\pm6.0 $~&~~0.001 ~&~~0.001\\
$\BR(K^+ \pi^-)$ ~&~~$19.4\pm0.6$ ~&~ $19.7\pm10  $~&~ $20.3\pm10  $~&~~0.001 ~&~~0.007\\
$\BR(K^0 \pi^0)$ ~&~~$9.9\pm0.6$  ~&~ $ 8.8\pm4.9 $~&~ $9.5\pm5.1 $~&~~0.046 ~&~~0.006\\
\hline
$\ACP(K^0 \pi^+)$ ~&~~$0.009\pm0.025$ ~&~~$ 0.0 \pm 0.0$  ~&~~ $ 0.0  \pm 0.0$~&~~0.13 ~&~~0.13\\
$\ACP(K^+ \pi^0)$ ~&~~$0.050\pm0.025$~&~~ $-0.017\pm0.068$~&~~$ 0.074\pm0.068$~&~~0.84 ~&~~0.11\\
$\ACP(K^+ \pi^-)$ ~&~~$-0.097\pm0.012$~&~~$-0.099\pm0.073$~&~~ $-0.11\pm0.075$~&~~0.001 ~&~~0.03\\
$\ACP(K^0 \pi^0)$ ~&~~$-0.14\pm0.11$~&~~  $-0.065\pm0.040$~&~~$-0.058\pm0.036$~&~~0.41 ~&~~0.50\\
\hline
$S_{K_S \pi^0}$ ~&~~ $0.38\pm0.19$~&~~ $0.74\pm0.08$~&~~ $0.74\pm0.07$        ~&~~3.1 ~&~~3.2\\
\hline
$\BR( \pi^+ \pi^0)$ ~~&~~$5.59^{+0.41}_{-0.40} $~&~~$4.03\pm2.53$~&~~$4.05\pm2.53$~&~~0.37 ~&~~0.36\\
$\BR( \pi^+ \pi^-)$ ~~&~~$5.16\pm0.22$~&~~$6.80\pm4.43$~&~~          $7.11\pm4.43$~&~~0.14 ~&~~0.19\\
$\BR( \pi^0 \pi^0)$ ~~&~~$1.31\pm0.21$~&~~$0.23\pm0.13$~&~~          $1.33\pm0.30$~&~~19 ~&~~ 0.002\\
\hline
$\ACP( \pi^+ \pi^0)$ ~~&~~$0.06\pm0.05$~~&~~$0.00\pm0.01$~~&~~         $0.055\pm0.018$~&~~1.6 ~&~~0.01\\
$\ACP( \pi^+ \pi^-)$ ~~&~~$0.38\pm0.07$~~&~~$0.17\pm0.10$~~&~~         $0.38\pm0.14$~&~~2.8 ~&~~0.0\\
$\ACP( \pi^0 \pi^0)$ ~~&~~$0.48^{+0.32}_{-0.31}$~~&~~$0.64\pm0.23$~~&~~$0.53\pm0.13$~&~~0.17 ~&~~0.024\\
\hline
$S_{\pi^+ \pi^-}$ ~~&~~ $-0.61\pm0.08$~~&~~ $ -0.55\pm0.44$~~&~~       $-0.55\pm0.42$~&~~0.021 ~&~~0.023\\
\hline
\end{tabular}
\end{table}
%%%%%%%%%%%%%%%%%%%%%%%%%%%%%%%%%%%%%%%%%%%%%%%%%%%%%%%%%%%%%%%%%%%%%%%%%%%
%%%%%%%%%%%%%%%%%%%%%%%%%%%%%%%%%%%%%%%%%%%%%%%%%%%%%%%%%%%%%%%%%%%%%%%%%%%%
%%%%%%%%%%%%%%%%%%%%%%%%  TABLE UPnum   %%%%%%%%%%%%%%%%%%%%%%%%%%%%%%%%%%%%%
\begin{table}[t]
\caption{
Numerical values of unparticle amplitudes for each decay mode. `Abs' represents
the magnitude of the amplitude in unit of $10^{-5}\GeV$. 'Arg' represents the
CP conserving phase of unparticle in radian unit.}
\label{tab:UPnum}
\begin{tabular}{c|cc||c|cc}
\hline
\hline
~~~~ decay mode ~~~&~ Abs ~&~ Arg ~&~~~ decay mode ~~~&~ Abs ~&~ Arg  \\
\hline
$K^0 \pi^+ $ ~~~&~~ 18.9 ~~~&~~  -1.6  ~&~ $\pi^+\pi^0 $ ~~~&~~ 0.6 ~~~&~~ 1.6    \\
$K^+ \pi^0 $ ~~~&~~ 10.8 ~~~&~~   1.6  ~&~ $\pi^+\pi^- $ ~~~&~~ 3.5 ~~~&~~ 1.6    \\
$K^+ \pi^- $ ~~~&~~  2.2 ~~~&~~  -1.6  ~&~ $\pi^0\pi^0 $ ~~~&~~ 9.9 ~~~&~~ 1.6    \\
$K^0 \pi^0 $ ~~~&~~  2.9 ~~~&~~   1.6  ~&~               ~~~&~~     ~~~&~~        \\
\hline
\end{tabular}
\end{table}
%%%%%%%%%%%%%%%%%%%%%%%%%%%%%%%%%%%%%%%%%%%%%%%%%%%%%%%%%%%%%%%%%%%%%%%%%%%
%%%%%%%%%%%%%%%%%%%%%%%%%%%%%%%%%%%%%%%%%%%%%%%%%%%%%%%%%%%%%%%%%%%%%%%%%%%%%%%%%%%%%%%%%%%%%%%%%%%%%%%%%%%%%%%%%%%%%%%%%%%%%%%%%%%%%%%%%%%%%%%%%%%%%%
Therefore, we define the $\chi^2$ to be
\begin{equation}
\chi^2 = \sum_{i=1}^n \frac{(t_i  - e_i)^2}{{\sigma^{exp}_i}^{~2}+{\sigma^{thr}_i}^2}\,.
\end{equation}
$e_i$ and $t_i$ denote the experimental data for $i$'th observable
and its theoretical prediction within unparticle contribution,
respectively. $\sigma^{exp}_i$ is the experimental error of $i$'th
observable, while $\sigma^{thr}_i$ is theoretical error propagated
from the errors of topological parameters obtained from PQCD. Since
there are $8$ free parameters involved in our analysis, the degree
of freedom (d.o.f) for the fitting is $(17-8)=9$. As for the angle
$\gamma$, we use the values of $\gamma = (63^{+15}_{-12})^\circ$
from the PDG 2006 \cite{Yao:2006px}. Consequently, by imposing the
mixing constraints of $C^{db}_{L(R)}$ and $C^{sb}_{L(R)}$ displayed
in Eq. (\ref{eq:CLCR}), we find the optimized values of unparticle
parameters as follows:
\begin{eqnarray}
C_L^{db} =  3.3\times 10^{-4}, ~ ~
C_R^{db} =  4.6\times 10^{-4}&,& ~ ~
C_L^{sb} =  7.6\times 10^{-4}, ~ ~
C_R^{sb} = 11.2\times 10^{-4}\,, \non \\
C_L^{uu} =  5.0,~~~~~ C_R^{uu} =  12.0&,&~~ C_L^{dd} =  4.2,~~~~~
C_R^{dd} =  11.2
\end{eqnarray}
with $\chi^2=4.6$, compared to $\chi^2=28.8$ without the unparticle
contributions. {}From above results, we see clearly $B_{d,s}-\bar
B_{s,d}$ mixings give strict constraints on $C^{db}_{L(R)}$ and
$C^{sb}_{L(R)}$ that lead to FCNCs at tree level. We compare the
experimental data to the theoretical predictions with and without
unparticle contributions in Table~\ref{tab:fit}, in detail. And
also, the numerical values of unparticle contributions are given in
Table~\ref{tab:UPnum} for comparing with the SM contribution given
in Table~\ref{tab:PQCD}. Strikingly, we can see that the large
experimental data of BR for $B_d \to \pi^0\pi^0$ can be quite well
accommodated with unparticle contributions. Moreover, some anomalous
observables of direct CPA of $B^+ \to K^+ \pi^0$ and $B_d \to \pi^+
\pi^-$ can be explained. However, the puzzle of mixing induced CPA
of $B_d \to K_S \pi^0$ could not be resolved well by unparticle
contributions. As many authors argued, sizable non-SM weak phase is
required in order to fit to the data of $S_{K_S \pi^0}$
\cite{Buras:2003dj}. Since the unparticle contributions do not carry
any extra weak phase, it turns out to be very hard to fit to the
data.

\section{Summary and Conclusions}

In summary, we have studied the effects of unparticle on $B_{u,d}$
decays with $B_{d,s}-\bar B_{d,s}$ mixing constraints. For
simplicity, we concentrate on vector unparticle and set the scale of
unparticle operator $\Lambda_{\UP}$ to be $1\,\TeV$. With the
lattice QCD results of JLQCD and (HP+JL)QCD on the non-perturbative
quantity of $f_{B_d}\sqrt{\hat B_{B_q}}$, we start to calculate the
SM predictions for $M^{q,SM}_{12}$. Accordingly, from the current
data we extract the available space for the new physics effects in
model independent way. When the unparticle effects are included to
$\Delta B=2$ processes, we find that the current experimental data
for $\Delta M_{d}$, $\Delta M_{s}$ and $\phi_d$ could give strict
constraints on unparticle parameters as well as scaling dimension
$d_{\UP}$. The (HP+JL)QCD case could not give a solution for
$d_{\UP}$, while all value of $d_{\UP}$ is possible in JLQCD case.
However, we see that more accurate SM prediction for the
$B^0_{q}-\bar B^0_{q}$ mixing amplitude in future would give strong
constraint on $d_{\UP}$. In order to understand whether the
unparticle effects could satisfy all measurements in exclusive
$B_{u,d}\to (\pi,\, K)\pi$ decays, we utilize the  minimum $\chi^2$
analysis to search for the solutions of free parameters.
Interestingly, after we fix $d_{\UP}=1.5$ for the specific
unparticle scaling dimension, we find that the unsolved problem of
large BR for $B_d \to \pi^0\pi^0$ could be explained excellently in
the framework of unparticle physics. Moreover the discrepancy
between the standard model estimation and data for the direct CPA of
$B^+ \to K^+ \pi^0$ and $B_d \to \pi^+\pi^-$ could be reconciled
very well. However, the puzzle of the mixing induced CPA of $B_d\to
K_S \pi^0$ could not be resolved well in unparticle physics.

\section*{Acknowledgments}
This work of C.H.C. is supported by the National Science Council of
R.O.C. under Grant No. NSC-95-2112-M-006-013-MY2. C.H.C. would like
to thank Dr. Shao-Long Chen for useful discussions. The work of C.S.K.
was supported in part by CHEP-SRC and in part by the KRF Grant
funded by the Korean Government (MOEHRD) No. KRF-2005-070-C00030.
The work of Y.W.Y. was supported by the KRF Grant funded by the
Korean Government (MOEHRD) No. KRF-2005-070-C00030. Y.W.Y thank Jon
Parry for his comment on mixing constraints.
\\
%\newpage

\end{document}